\documentclass[rnote]{aa} 
 
\usepackage{natbib}

\usepackage{graphicx}
\usepackage[varg]{txfonts}

\def\Feu  {\ion{Fe}{i}}
\def\Fed  {\ion{Fe}{ii}}

\def\Xu  {\ion{X}{i}}
\def\Xd  {\ion{X}{ii}}
\def\Teff  {$T_\mathrm{eff}$}
\def\logg  {$\log g$}

\def\vt    {$\rm v_{t}$}
\def\kms   {$\rm km\,s^{-1}$}
\def\st    {WISE\,J0725--2351}

\begin{document}

   \title{
Lithium abundance in a turnoff halo star on an extreme orbit.
\thanks{Based on observations obtained at the ESO Paranal Observatory, Chile Programmes 093.D-0127, PI: S. Geier
and 189.B-0925, PI: S. Trager.}
\fnmsep
\thanks{Table 2 with the line by line abundances of the elements, is available at the CDS via anonymous ftp to cdsarc.u-strasbg.fr (130.79.128.5) or via   
http://cdsarc.u-strasbg.fr/viz-bin/qcat?J/A+A/}
}
\author {
Spite M. 
\and
Spite F. 
\and 
Caffau E.
\and
Bonifacio P. 
}

\institute {
GEPI, Observatoire de Paris, PSL Research University, CNRS, Univ Paris Diderot, 
Sorbonne Paris Cit\'e, Observatoire  Place Jules Janssen, 92195 Meudon, France
}


\authorrunning{Spite et al.}

\titlerunning{Lithium abundance in a very high velocity halo turnoff star}

 
  \abstract
{The lithium abundance in  turnoff stars of the old population of our Galaxy is remarkably constant in the metallicity interval $\rm -2.8<[Fe/H] <-2.0$, defining a plateau. The Li abundance of these turnoff stars is clearly lower than the abundance predicted by the primordial nucleosynthesis in the frame of the standard Big Bang nucleosynthesis. Different scenarios have been proposed for explaining this discrepancy, along with the very low scatter of the lithium abundance around the plateau. 
}
{The recently identified very high velocity star, \st\  appears to belong to the old Galactic population, and  appears to be an extreme halo star on a bound, retrograde Galactic orbit. In this paper, we study the abundance ratios and, in particular the lithium abundance, in this star.   
}

{The available spectra (ESO-Very Large Telescope) are analyzed and the abundances of Li, C, Na, Mg,
Al, Si, Ca, Sc, Ti, Cr, Mn, Fe, Co, Ni, Sr and Ba are determined.
}
{The abundance ratios in  \st\  are those typical of old turnoff stars. The lithium abundance in this star ~is in close agreement with the lithium abundance found in the metal-poor turnoff stars located at moderate distance from the Sun. This high velocity star confirms, in an extreme case, that the very small scatter of the lithium plateau persists independent of the dynamic and kinematic properties of the stars.  
}
{}


\maketitle

%
\section{Introduction}
Recently, \citet{ScholzHH15} have studied a new high proper motion  metal-poor turnoff star,  with $\rm \mu=  267$ mas/yr  selected from a  high proper motion survey \citep{Luhman14},  based on observations with the Wide-field Infrared Survey Explorer  \citep[WISE; ][]{WrightEL10}. 

These authors obtained low- and medium-resolution spectra of this star:   WISE J072543.88--235119.7 
(hereafter \st). ~Their findings indicate that the star has a high radial velocity,
and thus a resulting Galactic rest frame velocity of about 460 \kms, showing that \st~ belongs to the extreme halo and crosses the Galactic disk on a retrograde bound orbit approaching the Sun at a distance of 400\,pc. 
 
\citet{ScholzHH15}  obtained spectra with the medium-resolution spectrograph X-shooter mounted on the ESO Very Large Telescope and deduced a first estimation of the atmospheric parameters of the star, \Teff = 6250 K, log g = 4.0,  [Fe/H ]=--2.0,  via  a comparison with the observed spectrum to a grid of theoretical spectra based on Kurucz models \citep{CastKur04,MunariSC05} computed with different temperatures, gravities, and metallicities. 
The low metallicity of the star shows that it pertains to the old population of the Galaxy. \citet{ScholzHH15}  also compared  the spectrum of \st~ with those of two classical Population II stars, HD\,84937 and HD\,140283, and found that the spectra are similar and that these stars should be also comparable in age. The age of HD\,140283 is estimated to be close to the age of the universe  following \citet{VandenBergBN14}.

The aim of the present paper is to check the lithium abundance in this extreme halo turnoff star.

\begin{figure}[h]
\resizebox{\hsize}{!}                   
{\includegraphics {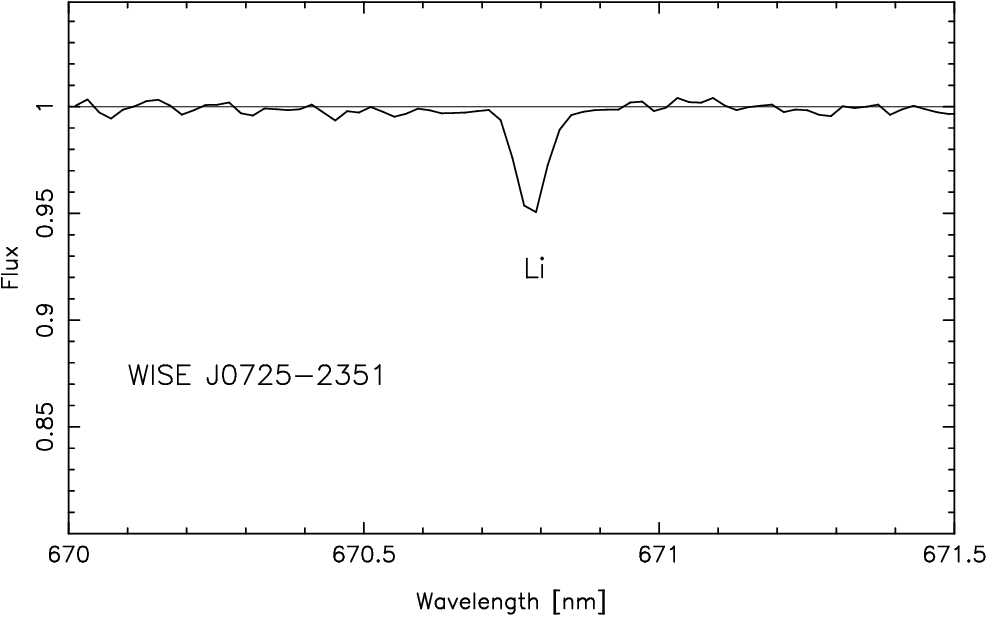}}
\caption[]{Observed profile of the lithium feature in \st.}
\label {Li}
\end{figure}

\section {Observations and analysis} 
The seven X-shooter spectra of \st, obtained during the ESO program ID 093.D-0127(A), were retrieved from the ESO archive, combined, and normalized. The resulting spectrum at 670\,nm, has  a mean resolution of about R=11000 and a S/N ratio per pixel of more than 500. 
It is thus possible to measure rather weak lines and, in particular, the lithium doublet is clearly visible (see Fig. \ref{Li}) and can be precisely measured. 

For the analysis of the star we used OSMARCS model atmospheres \citep{GustafssonEE08} 
and the {\tt turbospectrum} spectral synthesis code \citep{AlvarezP98,Plez12}.

\begin{figure}[ht]
\resizebox{\hsize}{!}                   
{\includegraphics {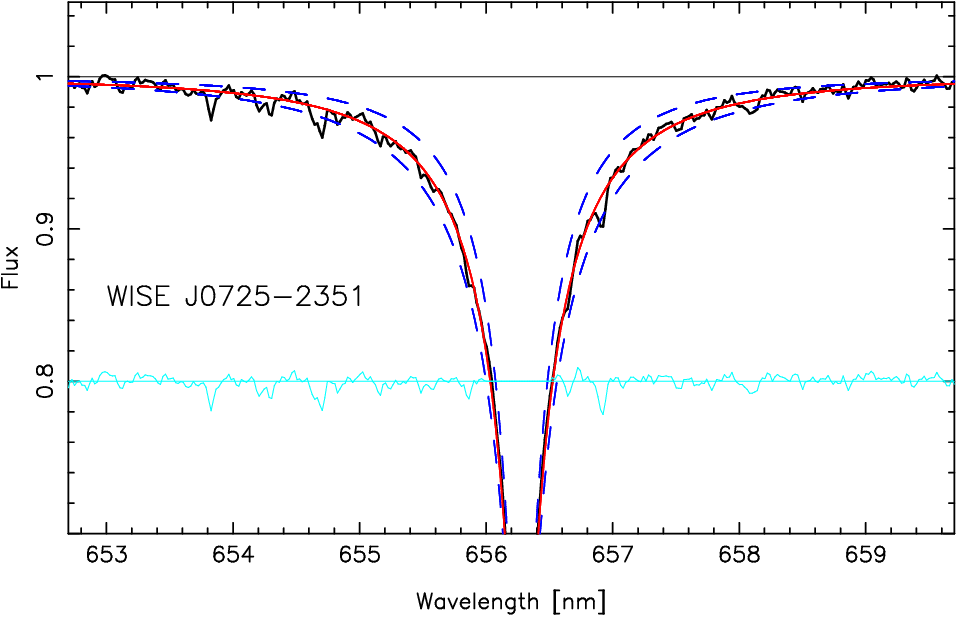}}
\caption[]{Observed profile of the $\rm H\alpha$ line in $\rm CD-23^{o}5447$ is compared to synthetic spectra computed with Teff ~= 5750 K and 6250 K(dashed blue lines) and Teff= 6050K (full red line: the best fit).}
\label {Halpha}
\end{figure}

In a first step we checked the temperature of the star by fitting the $\rm H_{\alpha}$ ~profile, which is a very good temperature indicator for turnoff stars \citep{Cayrel88,FuhrmannAG93,VVeerMeg96,BarklemSA02,BonMS07}.  
The theory of \citet{Barklem00} is used here, via {\tt turbospectrum,} to treat the hydrogen self-broadening, whereas 
\citep{ScholzHH15} use the
 grid of synthetic spectra of  \citet{MunariSC05}  based on the theory of \citet{AliGriem66} : this theory is known to lead to higher temperatures. We computed theoretical profiles of the $\rm H_{\alpha}$ line  with  three values of\Teff : 5750 K, 6000 K, and 6250 K. The best fit is obtained for a temperature of 6050 K; see Fig. \ref{Halpha}.\\
 
\begin{figure}[ht]
\resizebox{\hsize}{!}                   
{\includegraphics {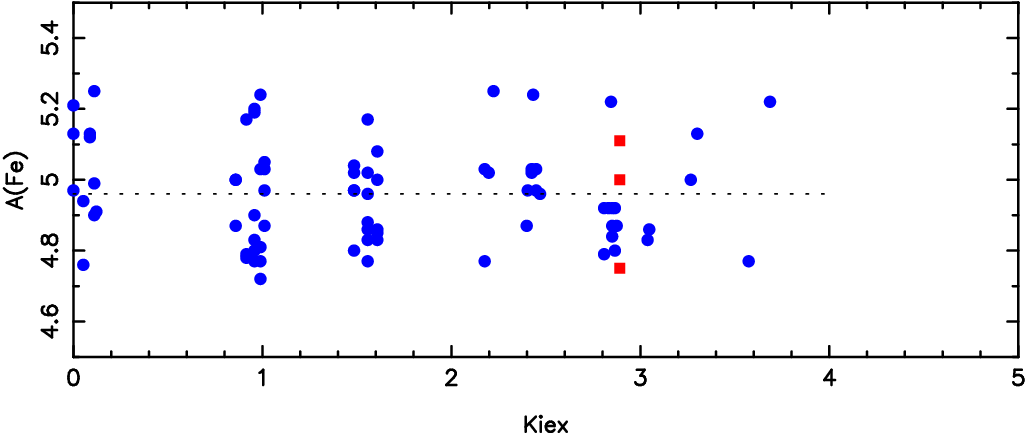}}
\caption[]{Comparison of the \Feu~ (blue dots) and \Fed~ (red squares) abundances vs. excitation potential for \st. The adopted model is \Teff=6050K \logg=4.2 and \vt=1.5\kms. There is no significant trend of the abundance with the excitation potential of the line.}
\label {Akiex}
\end{figure}

With this temperature, we performed a classical LTE analysis of \st ~using the code MyGIsFOS  \citep{SbordoneCB14}, as done by \citet{CaffauBS2013} in the frame of the ESO Large Programme TOPoS. The microturbulence velocity was taken equal to 1.5 \kms, and the surface gravity $\log g = 4.2$ ~was derived from the ionization equilibrium of iron. With these parameters, there is no significant trend in the \Feu~ abundance with the excitation potential of the line (see Fig.\,\ref{Akiex}). We estimated that that the total error in the adopted temperature is about 100K, the internal error in the estimation of the gravity is about 0.2\,dex, and the microturbulent velocity can be constrained within 0.2 \kms.\\
 
Following \citet{SitnovaZM15}, neglecting the NLTE effects in the determination of the gravity from the ionization equilibrium of iron  would lead, for metal-poor turnoff stars, to a shift in \logg~ up to 0.5 \,dex. 
The influence of this kind of shift on the abundance of the neutral elements and, in particular, on the lithium abundance ($<0.01$ \,dex for A(Li)) is negligible. The changes in [X/Fe] would be also very small if the Fe abundance deduced from the \Feu~ lines is used as a reference for the neutral species and the abundance deduced from the \Fed~ lines for the ionized species (model change would induce similar effects in the abundance of \Feu~ or \Fed,~ and \Xu~ and \Xd).
To allow an easier comparison to the previous works on similar stars, we prefer not to change the gravity of the model adopted.

The LTE abundances of the different elements are listed in  Table \ref{abund} (column 4), with the number of lines used for the computation (column 2)  and the adopted solar abundance in column 3; [X/Fe] is given in column 6 and the standard deviation in column 7.  The largest uncertainty in the abundance determination arises from the uncertainty in the temperature of the star. In general, however, model changes induce similar effects in the abundance of Fe and the other elements, and as a result they largely cancel out in the ratios [X/Fe] \citep{CayrelDS04,BonifacioSC09}. As a consequence the uncertainty in [X/Fe] is dominated by the random error in the lines.
Table 2 (available only at the CDS), lists the main parameters of the lines  (wavelength, excitation potential of the lower level, log gf) and the resulting abundance of the corresponding element.


The difference between the value of [Fe/H] in Table \ref{abund}  and the value obtained by \citet{ScholzHH15} is mainly due to the difference in the adopted temperatures. \st ~is located on the metal-rich edge of the interval of metallicity studied during the ESO Large Programme ``First Stars'' \citep{CayrelDS04,BonifacioSC09}.  In fact, the abundance ratios are very similar to those
found in the star CS\,31061-032 \citep{BonifacioSC09}, also on the metal-rich end.

\begin{figure}[ht]
\resizebox{\hsize}{!}                   
{\includegraphics {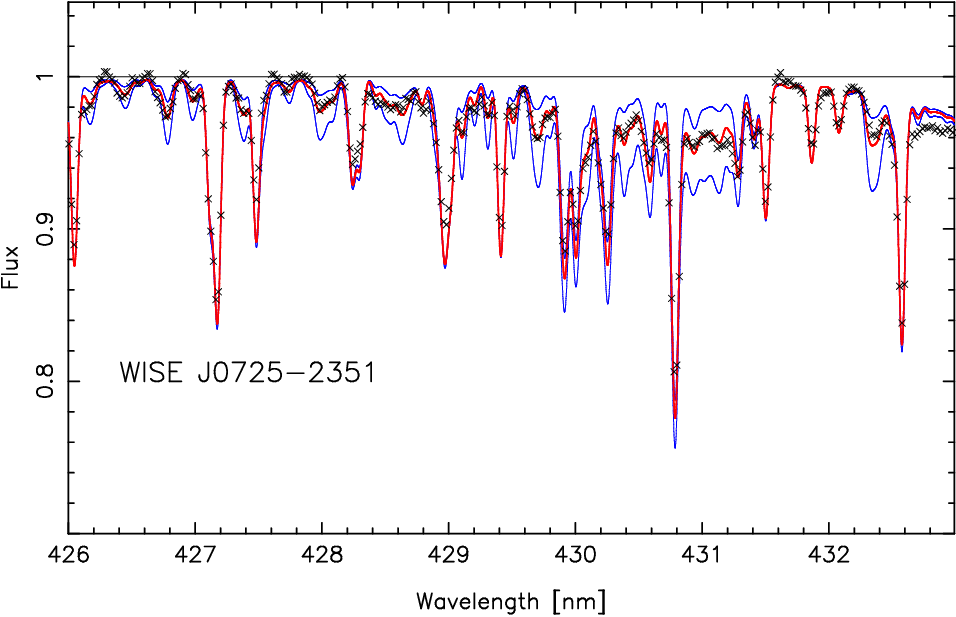}}
\caption[]{Observed profile of the CH band compared to synthetic spectra computed with A(C)=6.1 and 6.7 (thin blue lines). The best fit (thick red line) is obtained for A(C)=6.42.}
\label {CHband}
\end{figure}

As a sanity check of our analysis, we retrieved from the ESO archive an X-shooter spectrum of HD\,84937. It was observed on November 22 2013 with a 0\farcs{5} slit in the UVB arm and 0\farcs{7} in the VIS arm, yielding resolving power of 7900\relax\  in the UVB arm and 11000\relax\ in the VIS arm. The exposure time was 6s in both arms, providing S/N=79 at 402nm and S/N=71 at 670nm. We ran the code MyGIsFOS  on these spectra, with fixed effective temperature 6365\,K, log g = 4.0, and microturbulent velocity 1.6\,\kms: these atmospheric parameters were derived by \citet{Mashonkina08} from the analysis of high-resolution, high S/N spectra. With these parameters they derived [Fe/H] = --2.16, from the LTE analysis of ten \ion{Fe}{ii} lines.
The MyGIsFOS code, applied to the X-shooter spectra of our star, provides $\rm [Fe/H]_{II} = -2.23$ with a scatter of 0.13\,\,dex from three  \ion{Fe}{ii} lines, and  $\rm [Fe/H]_{I} = -2.20$ with a scatter of 0.22\,\,dex from 55 \ion{Fe}{i} lines.
In \st,~ the carbon abundance  (Fig. \ref{CHband}) is deduced from the profile of the CH G band. With [C/Fe]=0.49, \st~ has a carbon abundance very close to the mean value found for turnoff stars in \citet{BonifacioSC09}: $\rm\overline{[C/Fe]} = 0.45\pm 0.10$.

In Table \ref{abund}, the abundances of sodium and aluminum are deduced from the resonance lines and are strongly affected by NLTE effects. The NLTE correction for sodium is about $\rm -0.34\,dex$ \citep{AndrievskySK07}, and $\rm +0.70\,dex$ for aluminum \citep{AndrievskySK08}.
As a consequence, in the atmosphere of this star $\rm[Na/Fe]_{NLTE}=-0.27$ and $\rm[Al/Fe]_{NLTE}=+0.12$. These values are in good agreement with the mean values of [Na/Fe] and [Al/Fe] found in metal-poor stars at this metallicity \citep{AndrievskySK07,AndrievskySK08}. In particular, [Al/Fe] is significantly higher than the low value ($\rm[Al/Fe]_{NLTE}\sim -0.5$) found by \citet{GehrenSZ06} in a sample of halo stars, selected from their halo kinematics, at [Fe/H]=--2.5.

\begin{table}
\begin{center}    
\caption[]{
LTE abundances in \st~ obtained with \Teff=6050K \logg=4.2 and \vt=1.5\kms.
}
\label{abund}
\begin{tabular}{lccccccc}
\hline
Elem& N &$\rm A(X_{\odot})$&$\rm A(X_{\star})$&[X/H]&[X/Fe]&$\rm\sigma$\\
\hline
C(CH) &       & 8.50  &  6.42  & -2.08  &  0.49  &  -  \\        
Na I  &   2   & 6.30  &  3.80  & -2.50  &  0.07  &  -  \\
Mg I  &   4   & 7.54  &  5.56  & -1.98  &  0.59  & 0.22\\
Al I  &   2   & 6.47  &  3.32  & -3.15  & -0.58  & 0.14\\
Si I  &   1   & 7.52  &  4.98  & -2.54  & -0.03  &  -  \\
Ca I  &   5   & 6.33  &  4.09  & -2.24  &  0.33  & 0.15\\
Sc II &   1   & 3.10  &  0.72  & -2.38  &  0.19  &  -  \\
Ti II &   9   & 4.90  &  2.64  & -2.26  &  0.31  & 0.28\\
Cr I  &   4   & 5.64  &  2.78  & -2.86  & -0.29  & 0.23\\
Mn I  &   3   & 5.37  &  2.42  & -2.95  & -0.38  & 0.19\\
Fe I  &  77   & 7.52  &  4.95  & -2.57  &  0.00  & 0.20\\
Fe II &   3   & 7.52  &  4.95  & -2.57  &  0.00  & 0.26\\
Co I  &   1   & 4.92  &  2.83  & -2.09  &  0.48  &   - \\
Ni I  &   7   & 6.23  &  3.52  & -2.71  & -0.14  & 0.24\\
Sr II &   2   & 2.92  &  0.29  & -2.63  & -0.06  & 0.21\\
Ba II &   1   & 2.17  & -0.20  & -2.37  &  0.20  &  -  \\
\hline
\end{tabular}
\end{center}
\end{table}

\section {Lithium abundance} 
With the atmospheric model obtained in the previous section, we computed the lithium abundance. The equivalent width of the lithium feature is about 30\,m\AA, which leads to a  lithium abundance A(Li) = 2.17 (in a logarithmic scale with A(H)=12). We stress that the lithium abundance does not significantly depend on either  the gravity or the microturbulence velocity. Using the fitting formula of \citet{SbordoneBC10},   which provides the Li abundance taking both NLTE and granulation effects into account (a very small correction),  we finally find  A(Li) = 2.14.

\begin{figure}[h]
\resizebox{\hsize}{!}                   
{\includegraphics {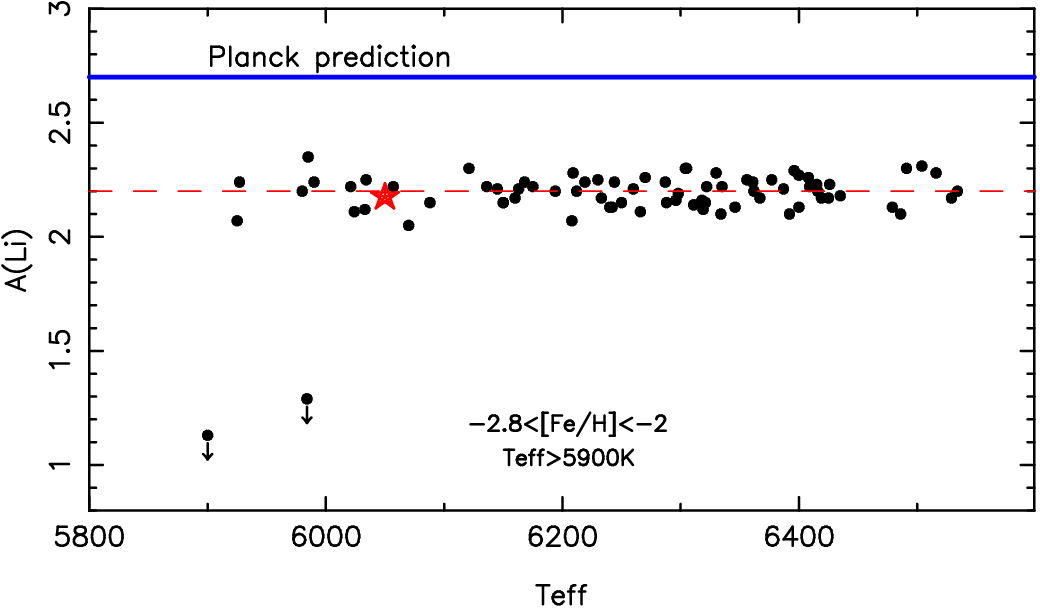}}
\caption[]{Lithium abundance vs. $\rm T_{eff}$ for stars selected for a temperature higher than 5900K (without deep convection) and without extreme deficiency $\rm -2.8<[Fe/H]<-2.0$.
Within these limits, the abundance of lithium is independent of the temperature and the
metallicity: $\rm A(Li)\approx 2.2$ (lithium plateau). The new high proper motion star (red star symbol) is located almost exactly at the mean value of the plateau.
}
\label {platLi}
\end{figure}

Lithium is a very fragile element destroyed as soon as the temperature exceeds 2.5 x $10^{6}$ K. If the convection zone is deep, as in cool stars, lithium is swept along to hot deep layers and is, little by little, destroyed. In Fig. \ref{platLi},  we  plotted the lithium abundance as a function of the effective temperature \Teff~ both for \st ~and for all the field galactic turnoff stars with the Li measurement available in the literature \citep{ChaPr05,ALN06,BonMS07,AokBB09,HosRG09,MelCR10,SbordoneBC10,SchK12}.
We only selected  the stars with temperatures higher than 5900K, and a metallicity between [Fe/H] = --2 and [Fe/H] = --2.8, as in \citet{SpiteSB12}.

The position of the extreme halo star \st ~in Fig. \ref{platLi} (very close to the mean value) confirms the very small scatter of the lithium abundance in the old galactic stars, in this interval of temperature and metallicity \citep{SpiteSB12}. The abundance A(Li) does not depend on the stellar kinematic or dynamic properties \citep[see also][]{BoesgaardSD05}.
At very low metallicity, when $\rm[Fe/H]<-2.8$, the lithium abundance in warm metal-poor dwarf stars becomes strongly scattered below the lithium plateau, this sudden melting down of the plateau at very low metallicity \citep{SbordoneBC10} has been discussed in \citet{BonifacioCS15} where different scenarios are proposed. However, \st\ is clearly in the metallicity regimewhere this effect is not seen.

\section {Conclusion}
The lithium abundance of \st~ has been carefully determined: A(Li)=2.17\relax\ in LTE and 
A(Li) = 2.14 \relax\ in 3D-NLTE. This recently discovered high velocity  star with a retrograde orbit belongs to the extreme halo according to \citet{ScholzHH15}.  This value confirms the low scatter of the lithium abundance (around the mean value (A(Li)=2.2) in the turnoff metal-poor stars with a metallicity higher than [Fe/H]=--2.8, independent of the kinematic or dynamic properties of the stars \citep[cf.][]{BoesgaardSD05}.

A lithium abundance similar to the mean value of the plateau has been also found in the turnoff stars of the galactic globular clusters, M\,92 and NGC\,6397 \citep{Bonifacio02,BonifacioPS02}; in a globular cluster $\rm\omega ~Cen$, considered a remnant of a dwarf galaxy \citep{MonacoBS10}; and in the globular cluster M\,54  \citep{MucciarelliSB14}, pertaining to the Sagittarius dSph galaxy. As a consequence, the mean lithium abundance in various old populations is significantly lower than the value predicted by the primordial nucleosynthesis after the Planck measurements \citep[see, e.g.,][]{Planck15,FieldsMS14,CyburtFO15}. Different scenarios concerning the formation of the early metal poor stars have been proposed to explain this difference   
\citep[e.g.,][]{MolaroBB12,FuBM15,BonifacioCS15}.

The chemical composition of \st~ resembles that of the average halo star in spite of its extreme kinematics. This suggests that the star has been formed ``in situ''. Nevertheless one can not rule out the possibility that it may have been accreted from a Milky Way satellite. In fact, at these low metallicities, some satellites have abundances that are remarkably similar to halo stars, in particular some stars show the typical $\alpha$/Fe enhancement, for example, in the dwarf galaxy Carina \citep{Koch08,Lemasle12,Venn12}, and in several of the ultra-faint dwarfs (Fran\c cois et al., in preparation).   
The orbit of \st ~ could suggest that this star was born outside of the Galaxy and was later accreted. In this case, it could be one more indication of the universality of  lithium abundance in the early universe.  

\begin {acknowledgements} 
This work has been supported by the ``Programme National de Physique Stellaire'' (CNRS-INSU), and it has made use of SIMBAD (CDS, Strasbourg). E.C. is grateful to the FONDATION MERAC for funding her fellowship.  
\end {acknowledgements}

\bibliographystyle{aa}
{}

\end{document}